\documentclass[final,3p,preprint, times,twocolumn]{elsarticle}
\usepackage{graphics}
\usepackage{graphicx}

\usepackage{amssymb}
\usepackage{amsthm}
\usepackage{amsmath}
\journal{Physics Letters B}

\begin{document}

\begin{frontmatter}

%% Title, authors and affiliationes

%% use the ead command for the email affiliation,
\title{Manifestation of the Berry phase in the atomic nucleus $^{213}$Pb}

\author[lnl]{J.J.~Valiente-Dob\'{o}n}

\author[lnl]{A.~Gottardo}

\author[infnmi]{G.~Benzoni}

\author[ific]{A.~Gadea}

\author[unipd,infnpd]{S.~Lunardi}

%\author{J.~Alc\'{a}ntara ~N\'{u}\~{n}ez}
%\affiliation{Universidade de Santiago de Compostela, Santiago de
%Compostela, E-175706, Spain}

\author[ific]{A.~Algora}

%\author{N.~Al-Dahan}
%\affiliation{Department of Physics, University of Surrey,
%Guildford, GU2 7XH, United Kingdom}

\author[lnl]{G.~de~Angelis}

%\author{Y.~Ayyad}
%\affiliation{Universidade de Santiago de Compostela, Santiago de
%Compostela, E-175706, Spain}

%\author{N.~Alkhomashi}
%\affiliation{KACST, Riyadh, 11442, Saudi Arabia}

%\author{P.R.P.~Allegro}
%\affiliation{Instituto de Fisica, Universidade de S\~{a}o Paulo,
%S\~{a}o Paulo, 05315-970, Brazil}

\author[infnpd]{D.~Bazzacco}

\author[santiago]{J.~Benlliure}

%\author{M.~Bowry}
%\affiliation{Department of Physics, University of Surrey,
%Guildford, GU2 7XH, United Kingdom}

\author[GSI]{P.~Boutachkov}

\author[infnmi,unimi]{A.~Bracco}

%\author{M.~Bunce}
%\affiliation{School of Computing, Engineering and Mathematics,
%University of Brighton, Brighton, BN2 4GJ, United Kingdom}

%\author{D. Brugnara}
%\affiliation{Istituto Nazionale di Fisica Nucleare, Laboratori
%Nazionali di Legnaro, Legnaro, 35020, Italy}

\author[brighton]{A.M.~Bruce}

\author[infnmi,unimi]{F.~Camera}

\author[vigo]{E.~Casarejos}

\author[GSI]{M.L.~Cort\'es}

\author[infnmi]{F.C.L.~Crespi}

\author[infnmi,unimi]{A.~Corsi}

%\author{A.M.~Denis Bacelar}
%\affiliation{School of Computing, Engineering and Mathematics,
%University of Brighton, Brighton, BN2 4GJ, United Kingdom}

%\author{A.Y.~Deo}
%\affiliation{Department of Physics, University of Surrey,
%Guildford, GU2 7XH, United Kingdom}

\author[GSI]{C.~Domingo-Pardo}

\author[salamanca]{M.~Doncel}

%\author{Zs.~Dombradi}
%\affiliation{Institute of Nuclear Research of the Hungarian
%Academy of Sciences, Debrecen, H-4001, Hungary}

\author[GSI]{T.~Engert}

%\author{K.~Eppinger}
%\affiliation{Physik Department, Technische Universit\"at
%M\"unchen, Garching, D-85748, Germany}

%\author{G.F.~Farrelly}
%\affiliation{Department of Physics, University of Surrey,
%Guildford, GU2 7XH, United Kingdom}

%\author{F.~Farinon}
%\affiliation{GSI Helmholtzzentrum f\"ur Schwerionenforschung,
%Darmstadt, D-64291, Germany}

\author[GSI]{H.~Geissel}

\author[GSI]{J.~Gerl}

%\author{A.~Goasduff}
%\affiliation{Istituto Nazionale di Fisica Nucleare, Laboratori
%Nazionali di Legnaro, Legnaro, 35020,
%Italy}

\author[GSI]{N.~Goel}

\author[GSI]{M.~G\'orska}

\author[cracow]{J.~Grebosz}

\author[GSI]{E.~Gregor}

\author[GSI]{T.~Habermann}

%\author{R.~Hoischen}
%\affiliation{GSI Helmholtzzentrum f\"ur Schwerionenforschung,
%Darmstadt, D-64291, Germany}\affiliation{Department of Physics,
%Lund University, Lund, S-22100, Sweden}

%\author{R.~Janik}
%\affiliation{Faculty of Mathematics and Physics, Comenius
%University, Bratislava, 84215, Slovakia}

\author[tum]{S.~Klupp}

\author[GSI]{I.~Kojouharov}

\author[GSI]{N.~Kurz}

\author[unipd,infnpd]{S.M.~Lenzi}

\author[infnmi,unimi]{S.~Leoni}

\author[delhi]{S.~Mandal}

\author[infnpd]{R.~Menegazzo}

\author[infnpd]{D.~Mengoni}

\author[infnmi]{B.~Million}

\author[infnmi]{A.I.~Morales}

\author[lnl]{D.R.~Napoli}

\author[GSI,koln]{F.~Naqvi}

\author[GSI]{C.~Nociforo}

\author[warsaw]{M.~Pf\"{u}tzner}

\author[GSI]{S.~Pietri}

\author[surrey]{Zs.~Podoly\'ak}

\author[GSI]{A.~Prochazka}

%\author{W.~Prokopowicz}
%\affiliation{GSI Helmholtzzentrum f\"ur Schwerionenforschung,
%Darmstadt, D-64291, Germany}

\author[infnpd]{F.~Recchia}

%\author{R.V.~Ribas}
%\affiliation{Instituto de Fisica, Universidade de S\~{a}o Paulo,
%S\~{a}o Paulo, 05315-970, Brazil}

%\author{M.W.~Reed}
%\affiliation{Department of Physics, University of Surrey,
%Guildford, GU2 7XH, United Kingdom}

\author[surrey]{P.H.~Regan}

\author[lund]{D.~Rudolph}

\author[lnl]{E.~Sahin}

\author[GSI]{H.~Schaffner}

\author[GSI]{A.~Sharma}

\author[comenius]{B.~Sitar}

\author[delhi]{D.~Siwal}

%\author{K.~Steiger}
%\affiliation{Physik Department, Technische Universit\"at
%M\"unchen, Garching, D-85748, Germany}

\author[comenius]{P.~Strmen}

%\author{T.P.D.~Swan}
%\affiliation{Department of Physics, University of Surrey,
%Guildford, GU2 7XH, United Kingdom}

\author[comenius]{I.~Szarka}

\author[infnpd]{C.A.~Ur}

\author[surrey,cern]{P.M.~Walker}

\author[GSI]{H.~Weick}

\author[infnmi]{O.~Wieland}

\author[GSI]{H-J.~Wollersheim}

%\author{I.~Zanon}
%\affiliation{Istituto Nazionale di Fisica Nucleare, Laboratori
%Nazionali di Legnaro, Legnaro, 35020, Italy}

%\author{F.~Nowacki}
%\affiliation{Universit\'{e} de Strasbourg, IPHC, 23 rue du Loess
%and CNRS, UMR7178, Strasbourg, 67037, France}

%\author{E.~Maglione} \affiliation{Departamento de F\'isica, Instituto Superior T\'ecnico,
%	Universidade de Lisboa, Lisbon, Portugal}
	
\author[ganil]{P.~Van~Isacker}

\address[lnl]{Istituto Nazionale di Fisica Nucleare, Laboratori
Nazionali di Legnaro, Legnaro, 35020, Italy}
	
\address[infnmi]{Istituto Nazionale di Fisica Nucleare, Sezione di
Milano, Milano, 20133, Italy}

\address[ific]{Instituto de F\'isica Corpuscular, CSIC-Universitat
de Val\`{e}ncia, Val\`{e}ncia, 46980, Spain}

\address[unipd]{Dipartimento di Fisica dell'Universit\`{a} degli
Studi di Padova, Padova, 35131, Italy}

\address[infnpd]{Istituto
Nazionale di Fisica Nucleare, Sezione di Padova, Padova, 35131, Italy}

\address[santiago]{IGFAE, Universidade de Santiago de Compostela, Santiago de Compostela, 15782, Spain}

\address[GSI]{GSI Helmholtzzentrum f\"ur Schwerionenforschung,
Darmstadt, 64291, Germany}

\address[unimi]{Dipartimento di Fisica
dell'Universit\`{a} degli Studi di Milano, Milano, 20133, Italy}

\address[brighton]{School of Computing, Engineering and Mathematics,
University of Brighton, Brighton, BN2 4GJ, United Kingdom}

\address[salamanca]{Grupo de F\'{i}sica Nuclear, Universidad de
Salamanca, Salamanca, 37008, Spain}

\address[vigo]{EEI, Universidade de Vigo, Vigo, 36310, Spain}

\address[cracow]{Niewodniczanski Institute of Nuclear Physics, Polish
Academy of Science, Krakow, 31-342, Poland}

\address[tum]{Physik Department, Technische Universit\"at
M\"unchen, Garching, 85748, Germany}

\address[delhi]{Department of Physics and Astrophysics, University of
Delhi, Delhi, 110007, India}

\address[koln]{Institut f\"ur
Kernphysik, Universit\"at zu K\"oln, K\"oln, 50937, Germany}

\address[warsaw]{Faculty of Physics, University of Warsaw, Warsaw,
00681, Poland}

\address[surrey]{Department of Physics, University of Surrey,
Guildford, GU2 7XH, United Kingdom}

\address[lund]{Department of Physics, Lund University, Lund,
	22100, Sweden}

\address[comenius]{Faculty of Mathematics and Physics, Comenius
University, Bratislava, 84215, Slovakia}

\address[cern]{CERN, Geneva,
1211, Switzerland}

\address[ganil]{Grand Acc\'el\'erateur National d'Ions Lourds, CEA/DRF-CNRS/IN2P3,  Caen, 14076, France}

\date{\today}% It is always \today, today,

\begin{abstract}
The neutron-rich $^{213}$Pb isotope was produced in the fragmentation  of a primary 1~GeV~$A$ $^{238}$U beam, separated in FRS in mass and atomic number, and then implanted for isomer decay $\gamma$-ray spectroscopy with the RISING setup at GSI. A newly observed isomer and its measured decay properties indicate that states in $^{213}$Pb are characterized by the seniority quantum number that counts the nucleons not in pairs coupled to angular momentum $J=0$. The conservation of seniority is a consequence of the Berry phase associated with particle-hole conjugation, which becomes gauge invariant and therefore observable in semi-magic nuclei where nucleons half-fill the valence shell. The $\gamma$-ray spectroscopic observables in $^{213}$Pb are thus found to be driven by two mechanisms, particle-hole conjugation and seniority conservation, which are intertwined through the Berry phase.
\end{abstract}


\end{frontmatter}

Observables in quantum mechanics are usually associated with the eigenvalue or expectation value of an operator. The major insight of Berry's paper~\cite{Berry84} is that a class of observables exists, which cannot be associated with any operator. If a Hamiltonian depends on a set of parameters $\{\xi\}$, then the phase change of an eigenstate of $\hat H(\xi)$ over a closed path in the parameter space is a gauge-invariant quantity, and as such observable. This simple result has far-reaching consequences and has found application in a wide variety of physical systems~\cite{Cohen:2019js}, a recent example being graphene~\cite{Dutreix19}. Berry's idea has also proven fruitful in nuclear physics with applications to pair transfer in superfluid rotating nuclei~\cite{Nikam87} and to the interplay between fast and slow degrees of freedom in the context of time-dependent Hartree-Fock-Bogoliubov theory~\cite{Bulgac90,Bulgac91} or the theory of collective motion~\cite{Girard90,Klein93,Kusnezov93}. Although of great theoretical interest, the observation of clear experimental signatures of these ideas remained elusive up to now. In this Letter we show that measurable properties of nuclei at mid-shell are strongly influenced by the Berry phase that connects particle-hole conjugation to seniority conservation.

Particle-hole conjugation is known since the early days of quantum mechanics (see, e.g., Chapter XII of Condon and Shortley~\cite{CS35}) and has found numerous seminal applications in atomic~\cite{Racah43} as well as nuclear~\cite{Bell59} physics. To this day it continues to inspire many branches of physics, a recent example being an application to the fractional quantum Hall effect~\cite{Haxton18}. Particle-hole conjugation transforms the problem of $n$ interacting fermions in a shell into one with $\Omega-n$ fermions, where $\Omega$ is the number of Pauli-allowed single-particle states. The $n$-fermion states correspond, up to a phase, to those for $\Omega-n$ fermions.  Together with an appropriate particle-hole transformation of the Hamiltonian, this leads to an identical interacting shell problem. 

Seniority $\upsilon$ is the number of particles not in pairs coupled to angular momentum $J=0$. As shown by Racah in the context of atomic physics~\cite{Racah43}, it is a conserved quantum number for $n$ identical particles, each with angular momentum $j$, interacting through a pairing force. This symmetry was later shown~\cite{Racah52,Schwartz54} to hold for a much wider class of interactions. A special case arises for {\em any} two-body interaction if the $j$ shell is half-filled with identical fermions, that is, if $n=(2j+1)/2$. 

The relation between particle-hole conjugation and seniority $\upsilon$ becomes clear by noting that an $n$-fermion state is transformed into a $(\Omega-n)$-fermion state times a phase. The phase is normally without any observable consequence {\em except} if either neutrons or protons half-fill a valence single-$j$ shell, in which case the original and transformed state are the same. This is an example of a gauge-invariant and hence observable phase, that is, of a Berry phase~\cite{Berry84}.

To obtain an understanding of the connection between seniority, particle-hole conjugation and the Berry phase, we start from a general particle-number-conserving, rotationally-invariant Hamiltonian with up to two-body interactions for a single-$j$ shell,
\begin{equation}
\hat H(E_0,\epsilon,\nu_\lambda)=
E_0+\epsilon\,\hat n-
{\textstyle\frac12}\sum_\lambda\nu_\lambda
(a^\dag a^\dag)^{(\lambda)}\cdot(\tilde a\tilde a)^{(\lambda)},
\label{e_ham}
\end{equation}
where $a^\dag_m$ creates a particle in the $j$ shell with projection $m$, $\tilde a_m\equiv(-)^{j+m}a_{-m}$ is the modified annihilation operator, $\hat n$ is the particle-number operator and $E_0$, $\epsilon$ and $\{\nu_\lambda,\lambda=0,2,\dots,2j-1\}$ are a constant energy, a single-particle energy and the two-body interaction matrix elements, respectively. The eigenstates of this Hamiltonian are characterized by particle number $n$, total angular momentum $J$, and its projection $M$. They depend on the parameters and can be denoted as follows:
\begin{equation}
|j^n\alpha JM;E_0,\epsilon,\nu_\lambda\rangle,
\label{e_eigen0}
\end{equation}
where $\alpha=1,2,\dots$ denotes the first, second, etc. energy eigenstate for a given $n$ and $J$. Since all Hamiltonians considered below are rotationally invariant, the projection $M$ is irrelevant and can be dropped.  

We introduce the following transformation from particle to quasi-particle:
\begin{equation}
\Gamma(\theta)a^\dag_m\Gamma(\theta)^\dag=
a^\dag_m \cos\theta+\tilde a_m\sin\theta.
\label{e_transfo}
\end{equation}
If this transformation is applied to the Hamiltonian~(\ref{e_ham}), eigenstates still carry $J$ as a quantum number but they no longer conserve particle number. The eigenstates of this transformed Hamiltonian belong to the Hilbert space spanned by the basis states $|j^n\upsilon J\rangle$. (In general, a label besides seniority $\upsilon$ is required to fully specify the state but it is not needed in the example below.)

The transformed Hamiltonian $\Gamma(\theta)\hat H(E_0,\epsilon,\nu_\lambda)\Gamma(\theta)^\dag$ has a more general form than the one in Eq.~(\ref{e_ham}) and the angle $\theta$ defines a certain path through the parameter space of the most general rotationally-invariant (but not necessarily particle-number conserving) Hamiltonian with zero, two and four particle creation and/or annihilation operators. Along this path, eigenstates of the transformed Hamiltonian pick up a phase factor $e^{i\phi(\theta)}$, known as the Berry phase~\cite{Berry84}.

We are only interested in the points of this path that correspond to a Hamiltonian that conserves particle number, in particular $\theta=0$ (no transformation) and $\theta={\frac12}\pi$ (particle-hole conjugation). It can be shown that
\begin{equation}
\Gamma({\textstyle\frac12}\pi)\hat H(E_0,\epsilon,\nu_\lambda)\Gamma({\textstyle\frac12}\pi)^\dag=
\hat H(E'_0,\epsilon',\nu_\lambda).
\label{e_transham}
\end{equation}
The constant term and single-particle energy are modified but the two-body interaction is invariant under the particle-hole transformation. The invariance of the interaction is crucial to the subsequent argument and, for example, a three-body interaction is {\em not} invariant under the particle-hole transformation and the derivation given below is {\em not} generally valid in that case.

The proof of the invariance of $\nu_\lambda$ can be found, for example, in Chapter 3 of Ref.~\cite{Lawson80}. Furthermore, the relation of $\Gamma({\frac12}\pi)$ to pairs of particles coupled to $\lambda=0$, and therefore to seniority, becomes apparent with its explicit representation~\cite{Mullerarnke73},  $\Gamma({\frac12}\pi)=e^{{\frac12}\pi(\hat S_+-\hat S_-)},$ in terms of $\lambda=0$ pair creation and annihilation operators $\hat S_+={\frac12}\sqrt{2j+1}(a^\dag_ja^\dag_j)^{(0)}_0$ and $\hat S_-=(\hat S_+)^\dag$. As a result each $\lambda=0$ pair in an $n$-particle state induces a minus sign under particle-hole conjugation such that the seniority basis transforms as follows:
\begin{equation}
\Gamma({\textstyle\frac12}\pi)|j^n\upsilon J\rangle=
(-)^{(n-\upsilon)/2}|j^{2j+1-n}\upsilon J\rangle.
\label{e_basis}
\end{equation}

Instead of following the evolution of the Berry phase along a continuous $\theta$-path, we may, following Pancharatnam~\cite{Pancharatnam56} (see also Resta~\cite{Resta00}), determine the phase directly for the discrete transformation with $\theta={\frac12}\pi$. Since $\Gamma({\textstyle\frac12}\pi)$ is a unitary operator,
$\Gamma({\textstyle\frac12}\pi)|j^n\alpha J;E_0,\epsilon,\nu_\lambda\rangle$ is an eigenstate of the transformed Hamiltonian
and we establish the identity
\begin{eqnarray}
\Gamma({\textstyle\frac12}\pi)|j^n\alpha J;E_0,\epsilon,\nu_\lambda\rangle&\!\!\!\!\!=\!\!\!\!\!&
\pm|j^{2j+1-n}\alpha J;E'_0,\epsilon',\nu_\lambda\rangle
\nonumber\\&\!\!\!\!\!=\!\!\!\!\!&
\pm|j^{2j+1-n}\alpha J;E_0,\epsilon,\nu_\lambda\}\rangle,
\label{e_eigend}
\end{eqnarray}
where the Berry phase $e^{i\phi({\frac12}\pi)}$ is either $+$ or $-$ because of Eq.~(\ref{e_basis}). The last equality holds since the constant and the single-particle energy do not affect the eigenfunctions of the Hamiltonian, only its eigenenergies. The sign $\pm$ is without any consequence except if the states on the left- and right-hand side of Eq.~(\ref{e_eigend}) are the same, which is the case for a half-filled shell, $n=2j+1-n$.

Consider as an example five particles in a $j=9/2$ shell. An eigenstate of the Hamiltonian~(\ref{e_ham}) can be expanded in the seniority basis,
\begin{equation}
|j^5\alpha J;E_0,\epsilon,\nu_\lambda\rangle=
\sum_{\upsilon=1,3,5}a_\upsilon|j^5\upsilon\,J\rangle,
\label{e_eigene}
\end{equation}
with coefficients $a_\upsilon$ that depend on the interaction matrix elements $\nu_\lambda$. Application of Eq.~(\ref{e_eigend}) with the $+$ sign, together with the relation~(\ref{e_basis}), yields
\begin{eqnarray}
&&a_1|j^5\upsilon=1\,J\rangle-
a_3|j^5\upsilon=3\,J\rangle+
a_5|j^5\upsilon=5\,J\rangle
\nonumber\\&\!\!\!\!\!=\!\!\!\!\!&
a_1|j^5\upsilon=1\,J\rangle+
a_3|j^5\upsilon=3\,J\rangle+
a_5|j^5\upsilon=5\,J\rangle.
\nonumber
\end{eqnarray}
Since the seniority basis is orthogonal, this implies that $a_3=0$ and that mixing can only occur between states with seniority $\upsilon=1$ and $\upsilon=5$. Likewise, application of Eq.~(\ref{e_eigend}) with the $-$ sign implies that $a_1=a_5=0$
and that the state has seniority $\upsilon=3$.

The argument can be generalized and leads to the result that at mid-shell a two-body interaction can only mix states that differ by four units of seniority, $\Delta\upsilon=\pm4$, confirming a well-known shell-model result~\cite{Lawson80}. While the Berry-phase mechanism explains total seniority conservation for $j\leq7/2$, for $j=9/2$ it explains the {\em partial} conservation of seniority~\cite{Escuderos06,Isacker08,Qian18} in self-conjugate semi-magic nuclei. It is an example of the general notion of partial dynamical symmetry~\cite{Alhassid92,Leviatan11}, which describes quantum-mechanical systems where some but not all eigenstates are solvable with a fixed structure that is independent of the parameters in the Hamiltonian.

The experimental results described in this Letter have their origin in the uniqueness of the FRS-RISING experimental complex~\cite{3,4,45, 5} and the UNILAC-SIS-18 accelerator facilities at GSI. The neutron-rich heavy nuclei beyond the $N=126$ shell closure were produced in the fragmentation of a 1~GeV/u $^{238}$U beam, with an intensity of around 1.5~$\times$~10$^9$ ions/spill. A beam extraction of $\sim$1~s was used, followed by a $\sim$2~s period without beam. The $^{238}$U ions were fragmented on a 2.5~g/cm$^2$ Be target, where a 223~mg/cm$^2$ Nb stripper was used, to improve the number of fully-stripped ions. The fragments produced in the reaction were separated according to their magnetic rigidity ($B\rho$) with the double-stage magnetic spectrometer FRS~\cite{3}. This experimental setup was described in detail in Refs.~\cite{lead,benzoni,Hg,morales,Bi,Tl,tesigottardo}. 
At the final focal plane of the FRS spectrometer, the fully-identified ions were slowed down in an Al degrader and eventually implanted in three layers of a double-sided silicon-strip detector (DSSSD) system~\cite{5,55}. The RISING $\gamma$ spectrometer consisted of 105~germanium crystals arranged in 15~clusters with 7~crystals each~\cite{4,45} surrounded by the implantation DSSSD system. 

\begin{figure}[t]
\begin{center}
\includegraphics[width=8.9cm]{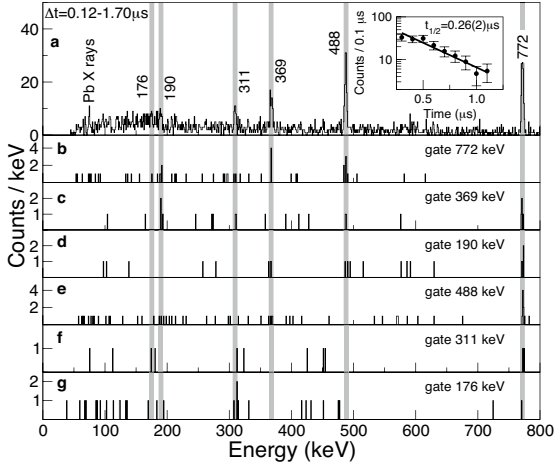}
\end{center}
\caption{(a) Gamma-ray spectrum showing the six $\gamma$-ray transitions assigned to the decay of the isomeric state in $^{213}$Pb. The spectrum is obtained by selecting the time window $\Delta t = 0.12$--$1.70~\mu$s after the implantation (referred to as delayed coincidence). The inset shows the time distribution and exponential fit for the sum of the transitions, yielding a half-life of $t_{1/2}=0.26(2)~\mu$s. \mbox{(b)--(g)} Prompt  $\gamma$-$\gamma$ coincidence spectra ($\Delta~t~\leq~100$~ns) between the six observed transitions following the isomer decay.} \label{213pbgamma}
\end{figure}
Figure~\ref{213pbgamma}~(a) shows the $\gamma$ spectrum in delayed coincidence with about 2100 fully-stripped $^{213}$Pb ions obtained by selecting a time window of $\Delta t = 0.12$--$1.70~\mu$s after implantation. Six $\gamma$ peaks with energies, in decreasing order, of 772, 488, 369, 311, 190,  and 176~keV are identified. The uncertainty on all energies is around $\pm1$~keV, except for the 176~keV transition, where it is $\pm2$~keV. The peak at 75~keV corresponds to K$_{\alpha}$ X-rays of Pb. The analysis of prompt $\gamma$-$\gamma$ coincidences, shown in Fig.~\ref{213pbgamma}~(b)--(g), indicates that the 772, 369, and 190~keV transitions are in mutual coincidence, while the 488~keV transition is only in coincidence with the 772~keV line. The 311 and 176~keV transitions, both in coincidence with the 772~keV, are also in mutual coincidence.  Furthermore, the sum of their energies, 311 and 176~keV, is compatible within uncertainties with the parallel 488~keV transition. %Importantly, the $\gamma$ coincidence spectra are very clean since background in these very selective spectra is almost non-existent. 
The $\gamma$ coincidence spectra present a low background; a 3-$\sigma$ significance has been considered for the $\gamma$-$\gamma$ coincidences.
Thus, as already shown~\cite{Hg} in $^{210}$Hg with half the statistics of $^{213}$Pb, even one or two coincidence counts, are significant. The yield of the 772~keV transition corresponds (after allowing for electron conversion) within uncertainties to the sum of the 369, 488, and 311 transitions. The lifetimes of the 369-, 772-, 488-, 311-keV $\gamma$ transitions are compatible among them. The 190- and 176-keV transitions do not have enough statistics for an individual lifetime determination. The inset in Fig.~\ref{213pbgamma}~(a) shows the exponential $\chi^2$ fit to the time decay curve for the sum of the $\gamma$ transitions, yielding a half-life of $t_{1/2}=0.26(2)~\mu$s for the isomeric state.  

\begin{figure}
\begin{center}
\includegraphics[width=8cm]{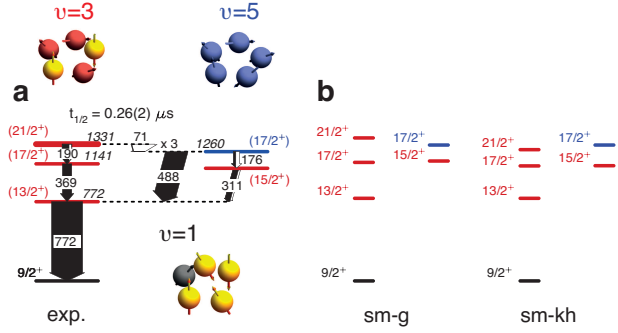}
\end{center}
\caption{(a)
Level scheme of $^{213}$Pb following the decay of the $21/2^+$ 0.26(2)~$\mu$s isomer deduced from the present data. The widths of the arrows are proportional to the relative $\gamma$-ray intensities but for display purposes the width of the 71~keV transition is reduced by a factor of three. The white part of the arrow is the internal conversion percentage of the transitions. Also shown is a schematic picture of the structure of the different states with seniority $\upsilon=1$ (black), $\upsilon=3$ (red), and $\upsilon=5$ (blue). The pairs of nucleons coupled to angular momentum $J=0$ are shown in yellow. (b) Theoretical predictions of a simple shell-model approach with five neutrons in $1g_{9/2}$ (sm-g) and of a large-scale shell-model calculation with the Kuo-Herling interaction (sm-kh), as described in the text.} 
\label{213pblevelscheme-3}
\end{figure}

A $J^{\pi}=21/2^+$ seniority isomer with a half-life of 42~ns, originating from the maximum coupling of three $1g_{9/2}$ neutrons, is known~\cite{Lane} in $^{211}$Pb. Its decay to the $9/2^+$ ground state is identified in a sequence of three $\gamma$-ray transitions, 137, 322, and 734~keV, feeding successively the $17/2^+$, $13/2^+$, and $9/2^+$ levels. An isomer with a longer half-life is expected in $^{213}$Pb as a consequence of the parabolic behavior of reduced transition probabilities $B(E2)$ predicted by the seniority scheme~\cite{Talmi93}. As discussed above, a similar sequence of coincident $\gamma$-rays 190, 369, and 772~keV is indeed observed in $^{213}$Pb. It is therefore associated with the cascade $21/2^+\rightarrow17/2^+\rightarrow13/2^+\rightarrow9/2^+$, from the isomer to the ground state. The 488, 311, and 176~keV transitions, in coincidence with the 772~keV transition, necessarily belong to a second decay branch of the $21/2^+$ isomer, absent in $^{211}$Pb. From energy arguments a 71~keV $\gamma$ ray, which cannot be observed experimentally because of its high internal conversion coefficient, is necessary. Thus, coincidences and intensities indicate that the $21/2^+$ isomer also decays via the two $\gamma$ sequences 71--488~keV and 71--176--311~keV. The ordering of the 71 and 488~keV transitions from the $21/2^+$ isomer to the $13/2^+$ level at 772~keV is determined by the half-life of the isomer, which is only compatible with an $E2$ $\gamma$-ray multipolarity for the 71~keV transition. Therefore, the isomer decays via a 71~keV $E2$ transition to a second $17/2^+_2$ level at 1260~keV followed by the 488~keV $E2$ transition to the $13/2^+$ level. This $17/2^+_2$ level also decays via a weaker branch (with around 15\% of the intensity) to the same $13/2^+$ level through the $\gamma$ sequence 176--311~keV, requiring an intermediate level at 949 or 1084~keV. The most straightforward solution is that, parallel to the 488~keV transition, the other two $\gamma$ rays are of $M1$ character, which suggests the spin and parity of the intermediate level to be $15/2^+$. It is worth noting that this second branch, decaying via the 71~keV transition, has a much higher intensity (around 80\% of the intensity) than the usual decay path $21/2^+\rightarrow17/2_1^+\rightarrow13/2^+\rightarrow9/2^+$, which carries the smallest fraction, around 20\%, of the intensity. Figure~\ref{213pblevelscheme-3}(a) shows the proposed level scheme for $^{213}$Pb, together with results of shell-model calculations. The calculations predict not only the expected cascade $21/2^+\rightarrow17/2_1^+\rightarrow13/2^+\rightarrow9/2^+$ but also two low-lying $17/2^+_2$ and $15/2^+$ states. These results agree with the experimentally inferred level scheme. The theoretical calculations favor the ordering of the 176 and 311~keV $\gamma$ lines shown in Fig.~\ref{213pblevelscheme-3}(a). The reduced $E2$ transition probabilities from the $21/2^+$ isomer to the two $17/2^+$ levels are determined as $B(E2;21/2^+\rightarrow17/2^+_1)=1.1(4)~e^2$fm$^4$ and $B(E2;21/2^+\rightarrow17/2^+_2)=32(5)~e^2$fm$^4$.

\begin{figure}
\begin{center}
\includegraphics[width=8cm]{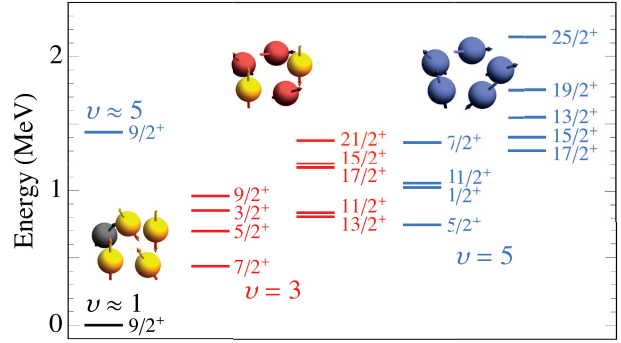}
\end{center}
\caption{Complete energy spectrum of five nucleons in the $1g_{9/2}$ shell. The spectrum applies to $^{213}$Pb and is calculated with two-body matrix elements derived by interpolation between $^{210}$Pb and $^{216}$Pb (sm-g). States can be of seniority $\upsilon=1$ (black), 3 (red), or 5 (blue), having two, one, or no pairs of nucleons coupled to angular momentum $J=0$ (yellow), respectively. All states conserve seniority, except for the two $9/2^+$ states with $\upsilon\approx1$ and $\upsilon\approx5$ shown on the left-hand side.} 
\label{f_g5}
\end{figure}

In the simplest shell-model approach the nucleus $^{213}$Pb
is described as five neutrons in the $1g_{9/2}$ shell.
The effective interaction between the neutrons can then be obtained either from the measured levels of $^{210}$Pb, from those of $^{216}$Pb, or by interpolation. The latter is used to calculate the (sm-g) spectrum of Fig.~\ref{213pblevelscheme-3}~(b). The complete $(1g_{9/2})^5$ spectrum is shown in Fig.~\ref{f_g5}. While the excitation energies do depend on the two-body matrix elements, the wave functions of most states in Fig.~\ref{f_g5} are fixed and have definite seniority.  For any two-body interaction, of all possible $(1g_{9/2})^5$ states only two can mix, namely those with $J^\pi=9/2^+$ and seniorities $\upsilon=1$ and $\upsilon=5$. All other states must have good seniority, either $\upsilon=3$ or $\upsilon=5$.

The partial conservation of seniority in mid-shell nuclei impacts on the electromagnetic decay. Matrix elements of the quadrupole operator between states of the same seniority are proportional to $2j+1-2n$ and vanish in mid-shell nuclei~\cite{Talmi93}, where $E2$ transitions therefore satisfy the selection rule $\Delta\upsilon=\pm2$. Specifically, the otherwise natural decay path of the $21/2^+$ level towards $17/2^+_1$ is forbidden while its decay to $17/2^+_2$ is allowed. This yields a qualitative explanation of the $21/2^+$ isomer decay: although from energy considerations the path via $17/2^+_2$ is around seven times less likely than that via $17/2^+_1$, experimentally it is found that the decay path via $17/2^+_2$ is approximately four times more probable. This observation is a consequence of seniority conservation. 

The results obtained with five neutrons in $1g_{9/2}$ will be altered by the presence of other shells in the valence space. In order to study the consequences of an increasing valence space, large-scale shell-model (LSSM) calculations were performed using the Kuo-Herling (KH) interaction~\cite{brown} in the full neutron valence space beyond $N=126$. Single-particle energies were extracted from the experimental spectrum of $^{209}$Pb for the $1g_{9/2}$, $0i_{11/2}$, $2d_{3/2}$, $2d_{5/2}$, $1g_{7/2}$, $3s_{1/2}$, and $0j_{15/2}$ neutron shells. The Hamiltonian was diagonalized using the $m$-scheme \textsc{antoine} and $J$-scheme \textsc{nathan} shell-model codes~\cite{antoine,nathan}. The latter allows one to obtain the seniority components of the calculated wave functions.

The results for the energies of the positive-parity states are shown in the (sm-kh) spectrum of Fig.~\ref{213pblevelscheme-3}(b) and found to be in good agreement with those of the simpler $1g_{9/2}$ approach and the observed energies. The expected cascade starting from the $21/2^+$ isomer proceeds through states with seniority $\upsilon=3$, decaying into the $9/2^+$ ground state with $\upsilon=1$. The  $B(E2;21/2^+\rightarrow17/2^+_1) =$~0.2~$e^2$fm$^4$, calculated with the standard neutron effective charge $e_\nu=0.5e$, is small and similar to the experimental value 1.1(4)~$e^2$fm$^4$. The overlap of the wave function of a LSSM state with a $\upsilon=3$ $(1g_{9/2})^5$ configuration is $0.84$ for $21/2^+$ and $0.83$ for $17/2^+_1$. A large $E2$ matrix element from $21/2^+$ to $17/2^+_2$ is also predicted. The $17/2^+_2$ LSSM state has an overlap with a $\upsilon=5$ $(1g_{9/2})^5$ configuration of $0.93$. The LSSM calculation therefore indicates the predominance of components with seniority $\upsilon=3$ and 5 in the $17/2^+_1$ and $17/2^+_2$ states, respectively. The calculated strength from the isomer to the $17/2^+_2$ level is $B(E2;21/2^+\rightarrow17/2^+_2)=65$~$e^2$fm$^4$, in line with the large value of 32(5)~$e^2$fm$^4$ observed experimentally. The relevant result is that the LSSM calculation predicts, as observed experimentally, substantially different $B(E2;21/2^+\rightarrow17/2^+_i)$ values, with the largest strength going to $17/2^+_2$. Of particular significance is also the finding that the LSSM calculation yields two $17/2^+$ states that are close in energy, 1098 and 1304~keV, but which remain pure in seniority, $\upsilon=3$ and 5, respectively. The shell-model calculation in the full valence space beyond $^{208}$Pb therefore confirms the approximate conservation of seniority, even if shells other than $1g_{9/2}$ are considered. Although the $1g_{9/2}$ shell is not isolated in energy, it is found to carry the dominant component of the wave function of low-energy states. 

We note that the KLS interaction and the truncation from Ref.~\cite{lead} cannot be used for testing seniority conservation since this approach includes only one-particle-one-hole excitations but no pair scattering from $1g_{9/2}$ to other shells.

Do other examples exist of semi-magic nuclei with valence nucleons that half-fill a single $j$ shell? For $j<7/2$, any state with a given total angular momentum $J$ is unique and conservation of seniority trivially follows from rotational invariance. The first non-trivial case occurs for four nucleons in the $0f_{7/2}$ shell, e.g. for $^{44}$Ca or $^{52}$Cr. Studies of such $0f_{7/2}$ nuclei reported considerable breaking of seniority in excited states as a result of mixing with the $1p_{3/2}$ shell~\cite{Auerbach67}. Concerning the next shell of interest with $j=9/2$, the best-studied nucleus is $^{95}$Rh with five protons in $0g_{9/2}$; it displays a decay of a $21/2^+$ isomer into two $17/2^+$ levels (see Ref.~\cite{jungclaus98} and references therein), similar to $^{213}$Pb. The isomer decay indicates substantially more seniority mixing in $^{95}$Rh, presumably as a result of neutron excitations across the $N=50$ closure~\cite{Mach17}. Another possible example is $^{213}$Fr with five protons in $0h_{9/2}$. A simple shell-model calculation predicts in this case that the $17/2^-_2$ level with seniority $\upsilon=5$ lies above the $21/2^-$ isomer, which therefore can only have a retarded decay to the $17/2^-_1$ level with seniority $\upsilon=3$. This finding agrees qualitatively with the isomeric character of the $21/2^-$ level and the non-observation of a $17/2^-_2$ level~\cite{Basunia07}. Finally, of $^{73}$Ni with five neutrons in $0g_{9/2}$, little is known at present.

In summary, our study of $^{213}$Pb shows that its $21/2^+$ isomer decays with asymmetric $E2$ transition probabilities to two $17/2^+$ states that are close in energy but of different nature. One state has two nucleons coupled to angular momentum $J=0$ (seniority $\upsilon=3$), while the other contains no $J=0$ pairs ($\upsilon=5$). The purity of seniority in the $17/2^+$ states follows from the self-conjugate character of $^{213}$Pb, where the particle-hole symmetry prevents $\Delta\upsilon=\pm2$ mixing. The property of seniority conservation in mid-shell nuclei is given a novel interpretation by means of an observable Berry phase. Further studies are required to deepen our understanding of the Berry phase and look for its additional consequences in nuclei. Exclusive cross-section measurements of one-nucleon transfer can be envisaged since pick-up and stripping reactions obey seniority selection rules. Finally, we note that the present experiment and discussion concerns a semi-magic nucleus and the conservation of seniority. The Berry-phase mechanism is more general, however, and an investigation of its observational consequences in nuclei with neutrons {\em and} protons at mid-shell is called for. Experiments of this kind will further enhance our understanding of symmetries in the atomic nucleus.\\

We thank F. Nowacki for providing the NATHAN and ANTOINE codes. The GSI accelerator staff are acknowledged. The authors acknowledge the support of the Italian Istituto Nazionale di Fisica Nucleare. This work was partially supported by the Ministry of Science, and Generalitat Valenciana, Spain, under the Grants SEV-2014-0398, FPA2017-84756-C4, PROMETEO/2019/005 and by the EU FEDER funds. The support of the UK STFC, of the Swedish Research Council under Contract No. 2008-4240 and No. 2016-3969 and of the DFG(EXC 153) is also acknowledged. The experimental activity has been partially supported by the EU under the FP6-Integrated Infrastructure Initiative EURONS, Contract No. RII3-CT-2004-506065 and FP7- Integrated Infrastructure Initiative ENSAR, Grant No. 262010.

\bibliographystyle{model1a-num-names}
\bibliography{bib}

\end{document}